%\vfill\eject
\beginsection
\shd{Proofs of theorems in quantum mechanics}

We begin with two propositions that allow us to transfer our considerations from
$\RR^{n+m}$ to the normal bundle $N\SS$. Let
$$
d(x,\SS)=\inf\{\|x-\ss\|:\ss\in\SS\}
$$
denote the distance to $\SS$ in $\RR^{n+m}$ and let
$$
\UU_\delta = \{x\in \RR^{n+m} : d(x,\SS)<\delta\}
$$
be the tubular neighbourhood of $\SS$ that is diffeomorphic to $N\SS_\delta$.
The first proposition shows that the time evolution
in $L^2(\RR^{n+m})$ under $H_\lambda$ is approximately the same for large $\lambda$ as the time 
evolution
in $L^2(\UU_\delta)$ under the same Hamiltonian, except with Dirichlet boundary conditions.

\beginpropositionlabel{putDir}
Suppose that $W,V\in C^\infty(\RR^{n+m})$ with $W\ge 0$ and $V$ bounded below.
Suppose $W(x)=0$ if and only if $x\in\SS$ and that $W(x) \ge w_0 > 0$ for large $x$. 

Suppose $\lambda \ge 1$, $\psi\in L^2(\RR^{n+m})$, $\|\psi\|=1$ and $\|H_\lambda\psi\|\le C_1\lambda^2$, where $H_\lambda=-\half\L+V+\lambda^4W$.
Then, given $\epsilon > 0$ there exists $C_2$ such that for all $t\in\RR$
\be{cutoffest}
\|F_{(d\ge\epsilon)}e^{-itH_\lambda}\psi\| \le C_2\lambda^{-1}.
\ee
Here $F_{(\cdot)}$ denotes multiplication by the characteristic function supported on the region indicated in the parentheses.

Define $H_\lambda^\delta$ be the operator in $L^2(\UU_\delta)$ given by $H_\lambda$ with Dirichlet boundary conditions on $\PA U_\delta$. Then for all $t\in [0,T]$ and $0<\epsilon<\delta$
\be{dirichletest}
\|F_{(d\le\epsilon)}e^{-itH_\lambda}\psi-e^{-itH_\lambda^\delta}F_{(d\le\epsilon)}\psi\| \le C_3\lambda^{-1/4}
\ee
Here $C_2$ depends only on $C_1$ and $\epsilon$ and $C_3$ depends only on $C_1$, $T$ and $\epsilon$.
\mcomment{actually $C_2$ also depends on $W$}
\endproposition
\beginremark The power $1/4$ in \rf{dirichletest} is not optimal. \endremark
\beginproof
By the assumption on $\psi$ and the Schwarz inequality
$$
\jap{\psi,H_\lambda\psi} \le C_1\lambda^2
$$
Without loss we may assume that $V\ge 0$,
so that
\be{gradbdds}\eqalign{
\half\|\nabla\psi\|^2 &\le C_1\lambda^2\cr
\jap{\psi,W\psi} &\le C_1\lambda^{-2}\cr
}\ee
It follows that
$$
C(\epsilon)\jap{\psi,F_{(d\ge\epsilon)}\psi} \le \jap{\psi,F_{(d\ge\epsilon)}W\psi} \le C_1\lambda^{-2}
$$
which proves \rf{cutoffest}, since $e^{-itH_\lambda}\psi$ satisfies the same hypotheses as $\psi$.

For $0<\epsilon_1\le\alpha$ we will need the estimate
\be{thingradest}
\|F_{(\epsilon_1\le d\le \alpha)}\nabla\psi\| \le C_4\lambda^{\half},
\ee
where $C_4$ depends only on $\alpha$, $\epsilon_1$ and $C_1$. To prove this, choose
a function $\chi\in C_0^\infty(\RR^{n+m})$, $0\le\chi\le 1$, which is $1$ in a
neighbourhood of $\{x:\epsilon_1\le d(x,\SS)\le\alpha\}$ and vanishes in a neighbourhood
of $\SS$. Then
$$
\|F_{(\epsilon_1\le d\le \alpha)}\nabla\psi\| = \|F_{(\epsilon_1\le d\le \alpha)}\nabla(\chi\psi)\| \le \|\nabla(\chi\psi)\|.
$$
The Schwarz inequality and integration by parts gives
$$
\|\nabla(\chi\psi)\| \le \|\L(\chi\psi)\|^\half\|\chi\psi\|^\half
$$
so that \rf{thingradest} follows from
\be{lapest}
\|\L(\chi\psi)\| \le C_5\lambda^2
\ee 
and \rf{cutoffest}. To prove \rf{lapest} let $p=-i\nabla$ and calculate, as forms on $C_0^\infty\times C_0^\infty$
\be{Hsquared}
H_\lambda^2 = {{1}\over{4}} |p|^4 + (V+\lambda^4W)^2 + \sum_jp_j(V+\lambda^4W)p_j - \half(\L V+\lambda^4\L W)
\ee
It follows from \rf{Hsquared} and the fact that $C_0^\infty$ is a core for $H_\lambda$ that $\chi\psi\in{\cal D}(H_\lambda)$ and
$$
\|\half p^2\chi\psi\|^2 \le \|H_\lambda(\chi\psi)\|^2 + C\lambda^4,
$$
or
$$
\half\|p^2\chi\psi\| \le \sqrt{C}\lambda^2 + \|H_\lambda\psi\| + \|[\half p^2, \chi]\psi\|.
$$
The last term can be bounded by \rf{gradbdds}, yielding \rf{lapest}.

\def\tchi{{\tilde\chi}}
Let $\tchi$ be a smooth function which satisfies $0\le \tchi\le F_{(d<\epsilon/2)}$ and $\tchi=1$ in a neighbourhood
of $\SS$. Because of \rf{cutoffest} (which holds at $t=0$) it is enough
to show
$$
\|e^{itH_\lambda^\delta}\tchi e^{-itH_\lambda}\psi - \tchi\psi\| \le C \lambda^{-1/4}
$$
for $t\in[0,T]$. Let
$$
\phi_{t,\lambda} = e^{itH_\lambda^\delta}\tchi e^{-itH_\lambda}\psi - \tchi\psi.
$$
Integrating the derivative, we obtain
$$\eqalign{
\phi_{t,\lambda} &= i \int_0^t e^{isH_\lambda^\delta}(H_\lambda^\delta\tchi -\tchi H_\lambda)e^{-isH_\lambda}\psi ds\cr
&=\int_0^t e^{isH_\lambda^\delta}(\nabla\tchi\cdot p-(i/2)\L\tchi)e^{-isH_\lambda}\psi ds,\cr
}$$
and thus
$$
\|\phi_{t,\lambda}\|^2 = \int_0^t\jap{e^{-isH_\lambda^\delta}\phi_{t,\lambda},(\nabla\tchi\cdot p-(i/2)\L\tchi)e^{-isH_\lambda}\psi}ds.
$$
\def\ttchi{{\tilde\tchi}}
Let $\ttchi=1$ on the support of $\nabla\tchi$ and $\ttchi=0$ in a neighbourhood of $\SS$. Then from \rf{thingradest}
$$\eqalign{
\|\phi_{t,\lambda}\|^2 &\le \int_0^t \|\ttchi e^{-isH_\lambda^\delta}\phi_{t,\lambda}\|
(\|\nabla\tchi\cdot pe^{-isH_\lambda}\psi\| + C)ds\cr
&\le C\lambda^\half\int_0^t\|\ttchi e^{-isH_\lambda^\delta}\phi_{t,\lambda}\|ds\cr
}$$
Now
$$\eqalign{
\jap{\phi_{t,\lambda}, H_\lambda^\delta\phi_{t,\lambda}}
&\le 2\jap{\tchi e^{-itH_\lambda}\psi,H_\lambda^\delta \tchi e^{-itH_\lambda}\psi} 
+ 2\jap{\tchi \psi,H_\lambda^\delta \tchi \psi}\cr
&=\jap{e^{-itH_\lambda}\psi,(H_\lambda\tchi^2+\tchi^2H_\lambda+(\nabla\tchi)^2)e^{-itH_\lambda}\psi} 
+\jap{\psi,(H_\lambda\tchi^2+\tchi^2H_\lambda+(\nabla\tchi)^2)\psi}\cr
\le C\lambda^2,
}$$
by the Schwarz inequality. Thus, following the proof of \rf{cutoffest},
$$
\|\ttchi e^{-isH_\lambda^\delta}\phi_{t,\lambda}\| \le C\lambda^{-1}
$$
so that
$$
\|\phi_{t,\lambda}\|^2  \le C\lambda^\half\lambda^{-1}
$$
which gives \rf{dirichletest}.
\endproof

Since the subset $\UU_\delta\subset\RR^{n+m}$ is diffeomorphic to $N\SS_\delta\subset N\SS$, we may
think of $H_\lambda^\delta=-\half\L + V + \lambda^4W$ as acting in $L^2(N\SS_\delta, d{\rm vol})$
with Dirichlet boundary conditions on $\PA N\SS_\delta$, where 
the volume form $d{\rm vol}$ and the Laplace operator $\L$ are computed using the pulled back
metric, and $V$ and $W$ are now the pull backs of the corresponding functions on $\UU_\delta$.
We may now extend the metric, and the potentials $V$ and $W$, from $N\SS_\delta$ to all of $N\SS$,
as explained in Section 4 above. Recall that the extended metric is assumed to be complete, that the extended
$V$ is bounded and that $W = \jap{n, A(\ss)n}$ on all of $N\SS$. We thus obtain an operator $H_\lambda$ 
acting in $L^2(N\SS,d{\rm vol})$. Since the extended metric
is complete,  $H_\lambda$ is essentially self-adjoint on $C_0^\infty$. Then it makes sense
to talk about $e^{-itH_\lambda}$. %\mcomment{reference?}

A proposition analogous to \thmrf{putDir} holds in this situation, allowing us to approximate
the evolution under $H_\lambda^\delta$ with an evolution under $H_\lambda$. For the purposes of this proposition, 
it does not matter how the extensions are made, as long as the conditions on the potentials hold, and the state $\psi$ that
we use for the comparison satisfies $\|H_\lambda\psi\|\le C\lambda^2$. Since the statement and
proof of this proposition are nearly identical to \thmrf{putDir} we omit them.

%For future reference, though,
%we just note the estimate
%\be{NScutoff}
%\|F(\|n\|\ge\epsilon)e^{-itH_\lambda}\psi\| \le C_2\lambda^{-1}.
%\ee 

Having justified the transfer of our considerations to $L^2(N\SS,\dvns)$, we now turn to the proof of \thmrf{quantum1}.

Before beginning, we need some quantum energy bounds. 
\beginlemmalabel{qenergy} Let $L_\lambda$ be as in \thmrf{quantum1} and $L_{0,\lambda}=H_B+\lambda^2H_O$.
Let $\Lsl$ denote either of these operators and $\Rsl = \left(\lambda^{-2}\Lsl + 1\right)^{-1}$.
Let $F_2 = F_{(|n|/\lambda < \epsilon)}$ be a smooth cutoff to the indicated region.
When $\epsilon < \delta$, this cutoff function is supported in the region of $N\SS$ where the metric is explicitly defined.
Let $\chi(\ss)$ be a cutoff with support in a single co-ordinate patch.
Then, for  small enough $\epsilon$ and large $\lambda$,
\be{ira8.1}
\|\jap{n}\Rsl^{1/2}\| + \|\chi F_2D_y\Rsl^{1/2}\| + \|\lambda^{-1}\chi F_2D_x\Rsl^{1/2}\| \le C
\ee
If $l$ is a non-negative integer and  $\alpha$, $\beta$ are multi-indices with $l+|\alpha|+|\beta|\le 2$, then
\be{ira8.2}
\|\chi F_2 \jap{n}^l (\lambda^{-1}D_x)^\alpha D_y^\beta\Rsl\| \le C.
\ee
In addition, if $l$ is a positive integer and $|\alpha|+|\beta|\le 2$, then
\be{ira8.3}
\|\chi F_2 \jap{n}^l (\lambda^{-1} D_x)^\alpha D_y^\beta \Rsl^{l+1}\| \le C.
\ee
Here  $\jap{n} = \sqrt{1+|n|^2}$.
\endlemma

\beginproof
Without loss of generality we can assume that $V\ge 1$. Set $f=\chi F_2$. Then
$f\in C_0^\infty$ with $0\le f\le 1$. Using \rf{ira7.4} we see that
$$
L_\lambda \ge \half (D-\PA k_\lambda)^*fG_\lambda^{-1} f (D-\PA k_\lambda) + {{\lambda^2}\over{2}}\sum_j \omega_j^2 y_j^2.
$$
In the region where $f>0$ we can use \rf{ira7.5} to obtain
$$
f\left[\matrix{I& -B\cr 0&I\cr}\right]^T\left[\matrix{I&0\cr 0&\lambda^2 I}\right]\left[\matrix{I& -B\cr 0&I\cr}\right]f
\le CfG_\lambda^{-1}f 
$$
Using $\lambda^{-2} R_\lambda^{1/2}(L_\lambda+\lambda^2)R_\lambda^{1/2} = 1$ we obtain
\be{ira8.4}
\|f D_y R_\lambda^{1/2}\| \le C
\ee
\be{ira8.5}
\lambda^{-1}\|f(D_x-BD_y-\PA_x k_\lambda + B\PA_y k_\lambda)R_\lambda^{1/2}\| \le C
\ee
\be{ira8.6}
\|
\jap{n}R_\lambda^{1/2}\| \le C.
\ee
On the support of $f$, $\PA_x k_\lambda - B\PA_y k_\lambda$ is bounded. Thus, using \rf{ira8.4} and
$\|B\|\le C|n|$ we obtain $\lambda^{-1}\|fD_x R_\lambda^{1/2}\| \le C$.
This proves \rf{ira8.1} for $R_\lambda$. The proof for $R_{0,\lambda}$ is similar. 

Define $U$ by $L_\lambda = \half D^* G_\lambda^{-1} D + U$. Then, using \rf{ira7.6} we calculate
$$\eqalign{
L_\lambda f^2 L_\lambda &= {{1}\over{4}}(f D^*G_\lambda^{-1}D)^*(f D^*G_\lambda^{-1}D) + D^*G_\lambda^{-1}f^2 U D + (Uf)^2 \cr
&\quad + \half D^* G_\lambda^{-1}[D,f^2 U] + \half [Uf^2,D^*]G_\lambda^{-1} D \cr
}$$
The last two terms above combine to give a multiplication operator given by a function which is easily shown to be
bounded below by
$$
-\tilde\chi^2\tilde F_2^2 (1+\lambda^2|y|^2)
$$
where $\tilde\chi$ and $\tilde F_2$ are like $\chi$ and $F_2$, with slightly expanded support. It follows that 
$$
{{\lambda^{-4}}\over{4}} \|f D^* G_\lambda^{-1}D R_\lambda\|^2 
+  \lambda^{-4}\|fG_\lambda^{-1/2}|U|^{1/2}DR_\lambda\|^2
+  \lambda^{-4}\|fUR_\lambda\|^2 \le 1 + \lambda^{-4}\|\tilde\chi \tilde F_2\jap{n}\lambda R_\lambda\|^2
$$
The right side is bounded by \rf{ira8.1}. 
From $\lambda^{-2}\|fUR_\lambda\|\le C$ we obtain $\|f\jap{n}^2 R_\lambda\|\le C$, which proves \rf{ira8.2} when
$l=2$.
From 
$$
\lambda^{-2}\|fG_\lambda^{-1/2}|U|^{1/2}D R_\lambda\|\le C
$$ 
we obtain
$$\lambda^{-1}\|f\jap{n}(D_x-BD_y)R_\lambda\|\le C$$ 
and
$$\|f\jap{n}D_yR_y\|\le C$$ 
which then gives
$$\|f\jap{n}\lambda^{-1}D_xR_\lambda\|\le C.$$
This proves \rf{ira8.2} when
$l=1$. Finally we consider the consequences of $\lambda^{-2}\|fD^*G_\lambda^{-1}DR_\lambda\|\le C$.
This is equivalent to 
$$\lambda^{-2}\|D^*G_\lambda^{-1}DfR_\lambda\|\le C$$ 
since the commutator term
can be bounded using \rf{ira8.1}. We thus must examine the operator $D^*G_\lambda^{-1}D$ acting on functions 
of compact support in $\RR^{n+m}$ contained in  a domain of the form 
$\Theta_\lambda = \{(x,y): |x|<r, |y| < \epsilon\lambda \}$ When we rescale $y\rarr\lambda y$ and $D_y\rarr\lambda^{-1}D_y$,
the operator $D^*G_\lambda^{-1}D$ goes over to an elliptic operator $E$ independent of $\lambda$ operating on functions of compact support
in a domain $\Theta = \{(x,y): |x|<r, |y| < \epsilon \}$. The smooth coefficients of the operator $E$ are bounded in $\Theta$. It follows that if $|\alpha|+|\beta|\le 2$
$$
\|D_x^\alpha D_y^\beta \psi\| \le C\|E\psi\|
$$
for $\psi$ with support in $\Theta$. When we scale back again this implies
$$
\lambda^{-2}\|D_x^\alpha (\lambda D_y)^\beta fR_\lambda\| \le C
$$
or
$$
\|(\lambda^{-1}D_x)^\alpha D_y^\beta fR_\lambda\| \le C.
$$
Again, the commutator term which arises from moving $f$ to the left can be bounded using \rf{ira8.1}.
This takes care of the case $l=0$ in \rf{ira8.2}.
We have thus proved
\rf{ira8.2} for $R_\lambda$. The proof for $R_{0,\lambda}$ is similar.

We now turn to \rf{ira8.3}. We give the proof for $R_\lambda$. The proof for $R_{0,\lambda}$ is similar. We first show that
\be{ira8.7}
\|f\jap{n}^l R_\lambda^l\|\le C.
\ee
We write $f=ff_1^l$ where $f_1$ has slightly larger support than $f$ and is of the form $h_1(x)h_2(|y|/\lambda)$.
Writing $f_1\jap{n}=g$, we have
$$\eqalign{
g^lR_\lambda^l &= g R_\lambda g^{l-1}R_\lambda^{l-1} + g[g^{l-1},R_\lambda]R_\lambda^{l-1}\cr
&= gR_\lambda g^{l-1}R_\lambda^{l-1} + g R_\lambda[\lambda^{-2}L_\lambda,g^{l-1}]R_\lambda^l\cr
&=gR_\lambda g^{l-1}R_\lambda^{l-1} + gR_\lambda\big(D_y^* J_1 \jap{n}^{l-1} + \lambda^{-1}(D_x-BD_y)^*J_2\jap{n}^{l-1}
+J_3\jap{n}^{l-1}\big)R_\lambda^l\cr
}$$
where $J_1$, $J_2$ and $J_3$ are bounded functions with support contained in ${\rm supp} f_1$. Thus, from \rf{ira8.1}
$$
\|g^lR_\lambda^l\| \le C \|g^{l-1} R_\lambda^{l-1}\| + C \|f_2\jap{n}^{l-1}R_\lambda^{l-1}\|
$$
where $f_2$ has slightly larger support than $f_1$. Thus \rf{ira8.7} follows inductively.

We now let $A_{\alpha,\beta}$ denote $(\lambda^{-1}D_x)^\alpha D_y^\beta$ and take $A= A_{\alpha,\beta}$
with $|\alpha|+|\beta|\le 2$. Then
$$
\|g^lA R_\lambda^{l+1}\| \le \|[A,g^l]R_\lambda^{l+1}\| + \|Af_2 g^l R_\lambda^{l+1}\|
$$
where $f_2$ has slightly larger support than $f_1$. We have
$$\eqalign{
\|Af_2g^l R_\lambda^{l+1}\| &\le \|Af_2 R_\lambda g^l R_\lambda^l\| + \|A f_2[g^l,R_\lambda]R_\lambda^l\|\cr
&\le \|Af_2 R_\lambda\|\cdot \|g^l R_\lambda^l\| + \|Af_2 R_\lambda\|\cdot\|[g^l,\lambda^{-2}L_\lambda]R_\lambda^{l+1}\|
}$$
and
$$
[A,g_l] = \sum_{|\gamma|+|\mu|\le 1} g_{\gamma,\mu,l-1}(\lambda^{-1}D_x)^\gamma D_y^\mu
$$
so that
$$
\|[A,g_l]R_\lambda^{l+1}\| \le \sum_{|\gamma|+|\mu|\le 1} \|g_{\gamma,\mu,l-1}A_{\gamma,\mu}R_\lambda^l\|.
$$
where $|g_{\gamma,\mu,l-1}|\le C (f_3\jap{n})^{l-1}$ and where $f_3$ has slightly larger support than $f_2$. 
Similarly
$$
[g^l,\lambda^{-2}L_\lambda] = \tilde J_1 \jap{n}^{l-1}D_y + \tilde J_2\jap{n}^{l-1}(\lambda^{-1}D_x) +  \tilde J_3\jap{n}^{l-1}
$$
where $\tilde J_1$, $\tilde J_2$ and $\tilde J_3$  are bounded functions with support contained in ${\rm supp} f_1$.
Thus
$$
\|[g^l,\lambda^{-2}L_\lambda]R_\lambda^{l+1}\| \le \sum_{|\gamma|+|\mu|\le 1}\|\tilde g_{\gamma,\mu,l-1}A_{\gamma,\mu}R_\lambda^l\|
$$
where $|\tilde g_{\gamma,\mu,l-1}|\le C (f_3\jap{n})^{l-1}$. Thus again using induction, the result \rf{ira8.3} follows.
\endproof

\beginproofof{\thmrf{quantum1}} Since
$$
\|e^{-itL_{0 \lambda}}\psi - e^{-itL_{\lambda}}\psi\|^2 = 2 \jap{\psi,\psi} - 2\Re\jap{\psi, e^{itL_{0 \lambda}}e^{-itL_{\lambda}}\psi}
$$
it suffices to show. 
\be{weaklim}
\lim_{\lambda\rarr\infty} \sup_{0\le t\le T}\left|\jap{\psi,e^{itL_{0 \lambda}}e^{-itL_\lambda}\psi} - \jap{\psi,\psi}\right| = 0
\ee
for a dense set of $\psi$ in $L^2(N\SS,\dvns)$. Let $\psi\in C_0^\infty$. Our goal is to show \rf{weaklim}.

As a first step, we insert an energy cutoff. Since 
$\|\Lsl \psi\| \le C\lambda^2$ we have
$$\eqalign{
\|F_{(\Lsl/\lambda^{2} \ge \mu)}\psi\| &= \|F_{(\Lsl/\lambda^{2} \ge \mu)}\Lsl^{-1}\| \cdot \|\Lsl\psi\|\cr
&\le C\mu^{-1}\cr
}$$
Set
$$
F_{\sharp 1} = F_{(\Lsl/\lambda^{2} \le \mu)}
$$
Then it suffices to show that for each fixed $\mu>0$
\be{ira8.9}
\lim_{\lambda\rarr\infty} \sup_{0\le t\le T}\left|\jap{F_{0 1}\psi,e^{itL_{0 \lambda}}e^{-itL_\lambda}F_1\psi} 
- \jap{F_{0 1}\psi,F_1\psi}\right| = 0.
\ee

We now need to show the quantum analogue of the fact in classical mechanics that the orbits stay in a bounded
region of phase space if we watch the system for a time $T<\infty$ which is independent of $\lambda$. 
Using energy considerations it follows from Lemma 8.2 that $\jap{n}$ and $D_y$ are bounded but only that $D_x$ cannot grow faster than $\lambda$.
We now seek a $\lambda$ independent bound, showing that up to a fixed time $T$, not too much energy can be transferred from normal to tangential modes.
In the quantum setting the statement
\be{ira8.10a}
\|F_{2} D_x \chi e^{-it\Lsl}F_{\sharp 1}\psi\| < C,
\ee
where $F_2$ is as in \thmrf{qenergy}, will suffice.

We will prove this estimate when $\Lsl = \Ll$, since the other case when $\Lsl = L_{0 \lambda}$is similar.
Let $\{\chi_k^2(\ss)\}$ be a partition of unity subordinate to a finite cover of co-ordinate charts. 
In other words, each $\chi_k^2$  is supported in a single co-ordinate chart, and $\sum_k\chi_k^2 = 1$. We may
assume that each $\chi_k$ is a smooth function only of $\ss$. Define
$$
Q = \sum_k \chi_k D_x^* G_\SS^{-1}(x)D_x \chi_k,
$$
where, in each term, $D_x$ and $x$ are defined in terms of the co-ordinates for 
the chart in which $\chi_k$ is supported. We now want to cut $Q$ off to the region where
we have explicit expressions for the metric, and then add a constant to regain
positivity. So let
$$
\bar Q = F_2 Q F_2 + 1
$$
Notice that $Q$ and $\bar Q$ commute with $F_{2}$, since
in local co-ordinates $F_2$ is a function of $y$ alone.
It is not difficult to show that both $Q$ and $\bar Q$
are essentially self-adjoint on $C_0^\infty(N\SS)$. Define
$$
q(t) = \jap{F_1\psi, e^{it\Ll}\bar Q e^{-it\Ll}F_1\psi}.
$$
Then \rf{ira8.1} follows from
$$
\sup\{q(t):t\in[0,T]\}\le C.
$$

We will prove a differential inequality as in the classical case. 
We will need further estimates to bound the terms which arise when
we compute $\dot q(t)$ and to prove an upper bound  for $q(0)$.

\beginlemmalabel{stuckinmiddle}
Suppose $F_1$ is a smooth cutoff in the energy $\lambda^{-2}L_\lambda$.
Then
$$
\left\|\Big(\jap{n}^l(\lambda^{-1}D_x)^\alpha D_y^\beta\Big) D_x^\gamma \chi_j F_2 F_1 \bar Q^{-1/2}\right\| \le C
$$
if $l+|\alpha|+|\beta| \le 2$ and $|\gamma|=1$.
\endlemma

\beginproof
We use the Helffer-Sj\"ostrand formula (see [D])
$$
F_1 = \int g(z) (R_\lambda - z)^{-1} dz\wedge d\bar z
$$
where  we may take $g\in C_0^\infty (\RR^2)$ with $|g(z)||\Im z|^{-N}\le C_N$ for any $N$. (We are
using the fact that $F_1(\lambda^{-2}L_\lambda) = \tilde F_1(R_\lambda)$ for
$\tilde F_1 \in C_0^\infty(0,2)$.
Let $A_1 = \jap{n}^\alpha(\lambda^{-1}D_x)^\beta D_y^\gamma \chi$ with $\chi\in C^\infty(\SS)$,
supported in the $j$th co-ordinate patch, $\chi\chi_1=\chi_1$, and let $F_{2,1}$ be a smooth
function of $|n|/\lambda$ with $F_{2,1}F_2 = F_2$. Then
$$
A_1 D_x^\gamma \chi_j F_2 F_1 \bar Q^{-1/2} = A_1 F_{2,1} F_1D_x^\gamma \chi_j F_2\bar Q^{-1/2} + 
A_1 F_{2,1} [D_x^\gamma \chi_j F_2, F_1]\bar Q^{-1/2}
$$
Using \rf{ira8.2}, the first term is bounded by a constant times 
$$
\|A_1F_{2,1}R_\lambda\|\cdot\|D_x^\gamma \chi_j F_2 \bar Q^{-1/2}\| \le C
$$
and it is thus sufficient to show
$$
\|R_\lambda^{-1}[D_x^\gamma \chi_j F_2, F_1]\| \le C.
$$
We compute from the Helffer-Sj\"ostrand formula
\be{ira8.10}
\|R_\lambda^{-1} [D_x^\gamma \chi_j F_2, F_1]\| \le C \|[D_x^\gamma \chi_j F_2, \lambda^{-2}L_\lambda]R_\lambda\|
\ee
For our present purposes we can write
$$
L_\lambda = (D_x-BD_y)^* E_0 (D_x-BD_y) + {{\lambda^2}\over{2}} (D_y^* D_y + \sum_j \omega^2 y_j^2) + E_0
$$
and we thus obtain
$$\displaylines{\quad
[D_x^\gamma \chi_j F_2, \lambda^{-2} L_\lambda] = \lambda^{-1} D_x^\gamma \chi_j(\nabla F_2\cdot D_y + D_y\cdot \nabla F_2)
\hfill\cr\hfill
+ \lambda^{-2}[D_x^\gamma \chi_j, (D_x-BD_y)^* E_0 (D_x-BD_y)]F_2 + \lambda^{-2} E_0
\quad\cr}$$
The first term gives a bounded contribution to \rf{ira8.10} by \thmrf{qenergy}. The second term can be
written
$$\eqalign{
&\Big(\lambda^{-1}(D_x-BD_y)^*E_0 \lambda^{-1}(D_x-BD_y) + D_y^* E_0 \lambda^{-1}(D_x-BD_y)\cr
& + \lambda^{-1}(D_x-BD_y)^*E_0 D_y +\lambda^{-2} E_0 (D_x-BD_y)\Big) \chi_jF_2\cr
& + \lambda^{-1} D_x^\gamma \Big( (\PA_x\chi_j)^T E_0 \lambda^{-1} (D_x-BD_y) + \lambda^{-1} (D_x-BD_y)^*E_0\PA_x\chi_j\Big) F_2\cr
}$$
and again this gives a bounded contribution to \rf{ira8.10} by \thmrf{qenergy}.\endproof

We now return to the proof of \thmrf{quantum1} and calculate
$$
\dot q(t) = i \jap{e^{-itL_\lambda} \psi, F_1 [L_\lambda, \bar Q] F_1 e^{-itL_\lambda}\psi}.
$$
Let $F_{1,1}$ be a $C_0^\infty$ function of $\lambda^{-2}L_\lambda$ with slightly larger support
than $F_1$, so that $F_1 F_{1,1} = F_1$. We will show that 
\be{ira8.11}
F_{1,1}[i L_\lambda, \bar Q] F_{1,1} \le C\bar Q
\ee
so that
$$
q(t) \le e^{Ct} q(0).
$$

First consider any term which arises when the cut-off $F_2 = F_{(|n|/\lambda < \epsilon)}$ is differentiated. 
The derivative $F_2'$ has support in a region of the form $\{(\ss,n):\lambda\epsilon_1 < |n| < \lambda \epsilon_2\}$
so that  $F_2'(\lambda/|n|)^l$ is bounded for any $l$. Thus
$F_2' = \big(F_2'(\lambda/|n|)^l\big)\lambda^{-l}|n|^l$ so that according to \thmrf{qenergy},
\rf{ira8.3}, such a term is bounded (and even decays faster that any inverse power of $\lambda$. Note that such a term occurring in the
commutator $[L_\lambda,\bar Q]$ appears alongside $D_x^\alpha D_y^\beta$ with $|\alpha| +|\beta| \le 3$
but because we have an $F_{1,1}$ on the left and another on the right, \rf{ira8.3} even allows $|\alpha|+|\beta|\le 4$ and we still obtain faster than any inverse power of $\lambda$ decay.) Since $\bar Q$ contains the constant $1$ such terms are harmless and we will
ignore them. Thus we are left with showing 
\be{ira8.12}
F_{1,1} F_2 [iL_\lambda, Q] F_2 F_{1,1} \le C \bar Q.
\ee
We write 
$$h_k = D_x^* G_\SS^{-1}(x) D_x$$
when the $x$ refers to the $k$th co-ordinate patch. Then
$$
\chi_k h_k\chi_k = \half \left(\chi_k^2 h_k + h_k \chi_k^2\right) + (\PA_x\chi_k)^TG_\SS^{-1}\PA_x\chi_k
$$
so that
$$\eqalign{
[L_\lambda,Q] &= \sum_k \left(\half [L_\lambda,\chi_k^2]h_k + \half h_k[L_\lambda,\chi_k^2] \right.\cr
&\hskip 20pt + [L_\lambda,m_k] + \left. \half \chi_k^2[L_\lambda,h_k] + \half[L_\lambda,h_k] \chi_k^2\right)\cr
}$$
where $m_k = (\PA_x\chi_k)^T G_\SS^{-1} \PA_x \chi_k$. We must make use of some cancellation which occurs above
so we write
$$
\sum_k \half [L_\lambda, \chi_k^2]h_k = \sum_{k,j} \half [L_\lambda,\chi^2_k](h_k-h_j)\chi_j^2
+\sum_{k,j} \half [L_\lambda,\chi_k^2]h_j \chi_j^2
$$
and note that the second term on the right vanishes because $\sum_k\chi_k^2 = 1$. Thus we obtain
$$\eqalign{
[L_\lambda,Q] &= \sum_{k,j} \half [L_\lambda,\chi_k^2](h_k-h_j)\chi_j^2 + \half\chi_j^2 (h_k-h_j)[L_\lambda,\chi_k^2]\cr
&\quad + [L_\lambda,{\cal M}] + \sum_k \half \chi_k^2 [L_\lambda, h_k] + \half [L_\lambda, h_k]\chi_k^2 \cr
}$$
where ${\cal M}= \sum_k m_k$. 

In the term $[L_\lambda,\chi_k^2](h_k-h_j)\chi_j^2$ we refer all operators to the $j$th co-ordinate patch. 
Thus
$$
h_k - h_j = \tilde D_x^* \tilde G_\SS^{-1}\tilde D_x-D_x^* G_\SS^{-1} D_x
$$
where $\sim$ refers to the $k$th co-ordinate system. We obtain (schematically)
$\tilde D_x = M^T D_x + \lambda E_1 D_y$ where $M\tilde G_\SS^{-1}M^T = G_\SS^{-1}$. Hence
$$
h_k-h_j = (\lambda E_1 D_y + E_0)D_x + \lambda^2 E_2 D_y D_y + \lambda E_1 D_y + E_0.
$$
After some calculation we find
$$\displaylines{\quad
\sum_{k,j} \half [L_\lambda,\chi_k^2](h_k-h_j)\chi_j^2 + \half\chi_j^2 (h_k-h_j)[L_\lambda,\chi_k^2]
\hfill\cr\hfill\lastdisplayline{ira8.13}{
\eqalign{
&=\sum_j \chi_j D_x^*(\lambda E_1 D_y + E_0) D_x \chi_j + \chi_j D_x^* (\lambda^2 E_2 D_y D_y + \lambda E_1 D_y + E_0)\cr
&\quad + \tilde\chi_j\left(D_y^*\lambda^3 E_3 D_y D_y + \lambda^2 E_2 D_y D_y + \lambda E_1 D_y + E_0\right)
}
\quad\cr}}$$
where $\tilde\chi_j\in C^\infty(\SS)$ with ${\rm supp} \tilde\chi_j$ contained in the $j$th co-ordinate patch. 
Noticing the presence of $F_2$ in \rf{ira8.12} and using \thmrf{stuckinmiddle} with $\alpha=0$ along with \rf{ira8.3} of
\thmrf{qenergy}, we see that the terms in \rf{ira8.13} give a contribution to the left side of \rf{ira8.12} which is bounded by $C\bar Q$.

We can re-expand ${\cal M}= {\cal M}(\ss)$ writing ${\cal M}=\sum_k{\cal M}\chi_k^2$ and then we find
$$
[L_\lambda,{\cal M}] = \sum_k \chi_k (D_x^* E_0 + \lambda E_1 D_y + E_0)
$$
which is readily handled by \thmrf{stuckinmiddle} and \rf{ira8.3} of \thmrf{qenergy}. We now expand the
terms involving $[L_\lambda, h_k]$. After some calculation we obtain
$$\displaylines{\quad
\sum_k \half\chi_k^2[L-\lambda, h_k] +\half [L_\lambda,h_k]\chi_k^2
\hfill\cr\hfill
\eqalign{
&= \sum_k \chi_k D_x^*\left(E_1 D_x + \lambda E_1 D_y + \lambda E_1 + E_0\right)D_x\chi_k\cr
&\quad + \sum_k \chi_k D_x^*\left((\lambda^2 E_2 + \lambda E_1)D_y D_y + \lambda E_1 D_y + \lambda E_1 + E_0\right)\cr
&\quad + \sum_k \chi_k\left((\lambda^2 E_2 + \lambda E_1)D_y D_y + (\lambda E_1 + E_0) D_y + \lambda E_1 + E_0 + \lambda^{-1} E_0\right)\cr
&\quad + \sum_k\left( \chi_k D_x^* E_1 D_x + \tilde\chi_k E_q D_x\right)\cr
}\quad}$$
where $\tilde\chi_k\in C^\infty(\SS)$ has support in the $k$th co-ordinate patch with $\tilde\chi_k\chi_k = \chi_k$. These terms are also easily 
handled with a combination of \thmrf{qenergy}, \rf{ira8.3} and \thmrf{stuckinmiddle}. This completes the proof of \rf{ira8.12} and
shows
$$
q(t) \le e^{Ct}q(0).
$$

Finally
$$
q(0) = \jap{F_1\psi,\bar Q F_1\psi}
$$
has $\lambda$ dependence and must be bounded uniformly in $\lambda$. But this follows from \thmrf{stuckinmiddle} (with $l=\alpha=\beta=0$)
and the fact that $\|\bar Q^{1/2}\psi\|^2 = \jap{\psi, \bar Q \psi} < \infty$, independently of $\lambda$.

We now return to \rf{ira8.9}. We introduce a stronger cutoff in the $n$ variable by restricting $|n|/\lambda^s < 1$ where
$s\in (0,1)$. Thus let $F_3 = F_{(|n|/\lambda^s < 1)}$ be a smooth cutoff the the indicated region. We note that
$$
\|(1-F_3)F_1\| \le \lambda^{-s}\|(1-F_1)\lambda^s/|n|\|\cdot \|\jap{n}F_1\| \le C\lambda^{-s}
$$
by \rf{ira8.1} of \thmrf{qenergy}. Thus it is sufficient to prove
$$
\lim_{\lambda\rarr\infty}\sup_{t\in[0,T]} 
\left|\jap{F_{0,1}\psi,e^{itL_{0,\lambda}}F_3 e^{-itL_\lambda}F_1\psi} - \jap{F_{0,1}\psi, F_3 F_1 \psi}\right|
=0
$$
By the fundamental theorem of calculus we obtain
$$\displaylines{\quad
\jap{F_{0,1}\psi,e^{itL_{0,\lambda}}F_3 e^{-itL_\lambda}F_1\psi} - \jap{F_{0,1}\psi, F_3 F_1 \psi}
\hfill\cr\hfill\lastdisplayline{ira8.14}{
= i\int_0^t \jap{F_{0,1}\psi, e^{isL_{0,\lambda}}
\left([L_{0,\lambda},F_3] + F_3(L_{0,\lambda}-L_\lambda)\right)e^{-isL_\lambda}F_1\psi}ds
\quad\cr}}
$$
The term $[L_{0,\lambda},F_3]$ contains derivatives of $F_3$ and thus by \thmrf{qenergy}, \rf{ira8.3} its
contribution to \rf{ira8.14} decays faster than any inverse power of $\lambda$ uniformly for $t\in [0,T]$.
According to \thmrf{Lexpression}, on the support of $F_3$ we have
$$
L_\lambda - L_{0,\lambda} = \sum_k \chi_k\Big( (D_x-BD_y)^* E_1(D_x-BD_y) + E_1\Big)\chi_k.
$$
Thus, aside from terms involving derivatives of $F_3$, which again can be handled by \thmrf{qenergy}, \rf{ira8.3}
we need only show that
$$
\lim_{\lambda\rarr\infty}\sup_{s\in [0,T]}
\left|\jap{F_{0,1} e^{isL_{0,\lambda}}\psi,\Big(F_2\chi_k(D_x-BD_y)^* F_3 E_1(D_x-BD_y)F_2\chi_k + \chi_k^2 F_3 E_1\Big)F_1e^{-isL_\lambda}\psi}\right|=0
$$
Now 
$$\|F_3 E_1 F_1\|\le C\lambda^{-1}\|\jap{n}F_1\|\le C\lambda^{-1}$$
so we need only bound the product
$$
\|(D_x-BD_y)\chi_k F_2 F_{0,1}e^{-sL_{0,\lambda}}\psi\|\cdot\|F_3 E_1\|\cdot \|(D_x-BD_y)\chi_k F_2 F_1 e^{-isL_\lambda}\psi\|.
$$
By \thmrf{qenergy}, \rf{ira8.2}
$$
\|BD_y\chi_k F_2 F_{\sharp, 1}\| \le C
$$
and by \rf{ira8.10a}
$$
\sum_{s\in [0, T]} \| D_x \chi_k F_2 F_{\sharp,1}e^{is\Lsl}\psi\|\le C.
$$
Finally
$$
\|F_3 E_1\| \le C\lambda^s/\lambda = C \lambda^{s-1},
$$
which proves \rf{ira8.9} and thus completes the proof of the theorem. \endproof

\endsection